\documentclass[reprint,amsmath,amssymb,aps,prx, superscriptaddress,floatfix]{revtex4-1}

\usepackage{hyperref}
\usepackage{graphicx}
\usepackage{mathtools}
\usepackage{multirow}
\usepackage{array}
\usepackage{tikz}
\usepackage[normalem]{ulem}

\usepackage{hyperref}
\hypersetup{
colorlinks=true,final=true,
        linkcolor=blue,
        citecolor=blue,
        filecolor=blue,
        urlcolor=blue,
}

\DeclareMathOperator{\im}{Im}

\begin{document}

\title{Electron vs. hole doping in infinite-layer nickelates: electronic structure, magnetism and correlations}

\author{Ezra Day-Roberts}
\affiliation{Department of Physics, Arizona State University, Tempe, AZ 85287, USA}
\author{Fabio Bernardini}
\affiliation{Dipartimento di Fisica, Università di Cagliari, IT-09042 Monserrato, Italy}
\author{Harrison LaBollita}
\affiliation{Center for Computational Quantum Physics, Flatiron Institute, New York, New York 10010, USA}
\author{Yi-Feng Zhao}
\affiliation{Department of Physics, Arizona State University, Tempe, AZ 85287, USA}
\author{Andres Cano}
\affiliation{Univ. Grenoble Alpes, CNRS, Grenoble INP, Institut Néel, 25 Rue des Martyrs, 38042 Grenoble, France}
\author{Antia S. Botana}
\affiliation{Department of Physics, Arizona State University, Tempe, AZ 85287, USA}

\begin{abstract}

The observation of superconductivity in undoped infinite-layer nickelates $R$NiO$_2$ ($R$ = rare earth) challenges our current understanding and calls for a re-examination of the underlying electronic structure of this family of materials. In this context, it is particularly important to extend the investigation of $R$NiO$_2$ compounds from the intensively studied hole-doped regime to the almost unexplored electron-doped one. Here, we use a combination of density-functional theory and dynamical mean-field theory to study the evolution of the electronic structure of infinite-layer nickelates in these two doping regimes. 
We find a striking asymmetry in the self-doping of the Ni-$d_{x^2-y^2}$ band due to the $R(5d)$ states: 
while this effect is strongly suppressed upon hole doping, electron doping instead leads to an increase in the size of the $R(5d)$ electron pockets, but without effectively hole-doping 
the Ni-$d_{x^2-y^2}$ band. 
This asymmetry has an important impact on the magnetic response as antiferromagnetism is rapidly suppressed upon hole doping, whereas it remains the ground state upon electron doping. Despite these differences, electronic correlations on both sides of the phase diagram are dominated by the Ni $d_{x^2-y^2}$ orbital, suggesting that a single-band description may be appropriate for infinite-layer nickelates in both the electron- and hole-doped regimes.

\end{abstract}
\maketitle

\section{Introduction}

\begin{figure}[h!]
    \centering
   \includegraphics[width=0.92\columnwidth]{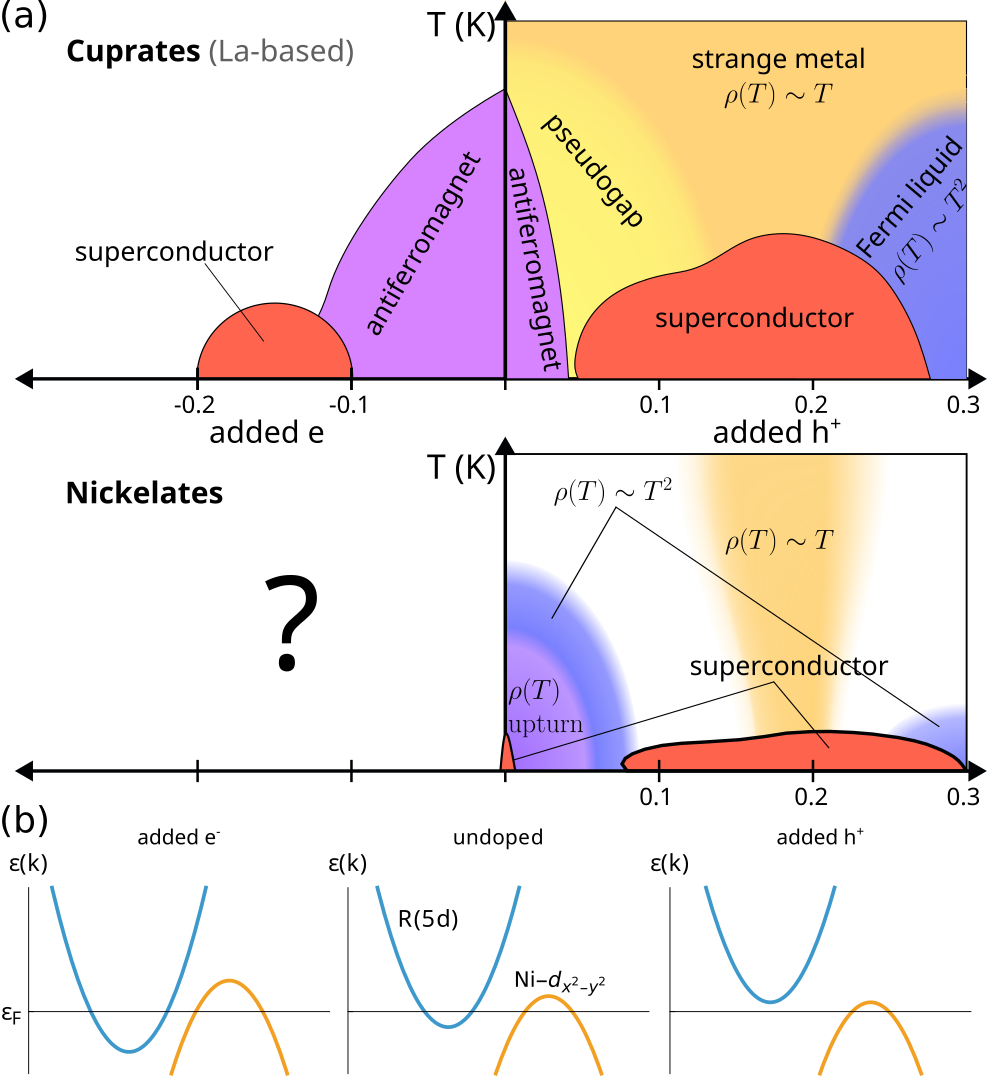}
    \caption{(a) Electron- and hole-doped phase diagrams for superconducting cuprates (top) and nickelates (bottom) where distinct electronic phases are denoted in different colors. Cuprate diagram adapted from Ref. \cite{Armitage2010}. Nickelate phase diagram adapted from transport data in Refs. \cite{hsu2024,Wang2024}. 
    (b) Schematic of the near Fermi level band structure for electron-doped vs. hole-doped infinite-layer nickelates.}
    \label{fig:schematics}
\end{figure}

The discovery of superconductivity in infinite-layer nickelates $R$NiO$_2$ ($R$= rare-earth) 
opened new avenues in the search for high-critical temperature (T$_c$) superconductors \cite{Li2019}. These materials exhibit remarkable similarities to the cuprates displaying two-dimensional transition-metal oxygen square planes and a nominal $d^{9}$ electron filling  ~\cite{botana2020,Karp2020:112,olevano20,lechermann2020late,lechermann2020multi,lechermann2022,labollita2021,Bernardini2021,Wissel2022,worm2021correlations,kitatani2020,been2021:prx,Leonov2020,kang2023}. 
Mapping the phenomenology of the infinite-layer nickelates to the hole-doped cuprate phase diagram has proven fruitful for comparing the physics of these two material families (see Fig.~\ref{fig:schematics}). While early efforts found T$_c$s limited to $\sim$ 15 K upon hole doping in $R$NiO$_2$ \cite{Li2019,Osada2020,Osada2021nickelate,Zeng2021superconductivity}, increased transition temperatures have been recently achieved upon chemical~\cite{Chow2025} and hydrostatic pressure \cite{Lee2026_NiO2_membrane_pressure,Wang2022}. Despite several structural and electronic parallels, infinite-layer nickelates exhibit unique features absent in cuprates, such as additional bands of $R(5d)$ character crossing the Fermi level and a reduced hybridization with the oxygen ions. While the role of these $R(5d)$ bands in the normal and superconducting state of infinite-layer nickelates remains an open question \cite{ES_112, arita, Liu2020, Thomale_PRB2020, Choi2020, Ryee2020, olevano20,Leonov2020,lechermann2020late,lechermann2020multi, Hu2019, Sakakibara, werner2020, eff_ham, Zhang2020, kitatani2020}, several works have pointed out that the $R(5d)$ states simply act as a charge reservoir, self-doping the Ni-$d_{x^2-y^2}$ band~\cite{labollita2023:conductivity, kitatani2020}. 

Recent experimental reports have challenged our basic understanding of infinite-layer nickelates, as superconductivity has now been observed in stoichiometric $R$NiO$_2$ across several rare-earth elements ($R$ = La-Nd) \cite{Osada2021nickelate,Sahib2025,Parzyck2025}. 
This observation is consistent with the self-doping effect introduced by the $R(5d)$ ``spectator" bands, which dope the Ni-$d_{x^{2}-y^{2}}$ orbital away from its nominal $d^{9}$ configuration and may as such promote superconductivity. 
These emerging experimental results also highlight the need to expand theoretical investigations to the electron-doped side of the infinite-layer nickelate phase diagram. 
This extension (see Fig. \ref{fig:schematics}) can provide critical insights into the connections between the nickelates and cuprates, as well as into the nature of their underlying superconductivity instabilities. Importantly, electron doping of the parent $R$NiO$_3$ phases with Ce$^{4+}$ and Th$^{4+}$ has been reported~\cite{song2023,ALONSO1995146}. If these phases can be topotatically reduced while maintaining the $4+$ oxidation state of the dopant, the corresponding electron-doped square-planar infinite-layer nickelates could be accessed.

To address these open questions, we explore the normal state electronic structure of the infinite-layer nickelate LaNiO$_2$ upon electron-doping in comparison to hole-doping using a combination of density-functional theory (DFT) and dynamical mean-field theory (DMFT). Our electronic structure calculations reveal a clear asymmetry between the electron and hole-doped sides of the phase diagram in terms of their magnetic properties and electronic structure. Our findings establish  electron-doping as an interesting avenue to pursue experimentally in $R$NiO$_2$ compounds, specifically in the context of sharpening the analogies to the cuprates.

\begin{figure*}
    \centering
    \includegraphics[width=0.9  \linewidth]{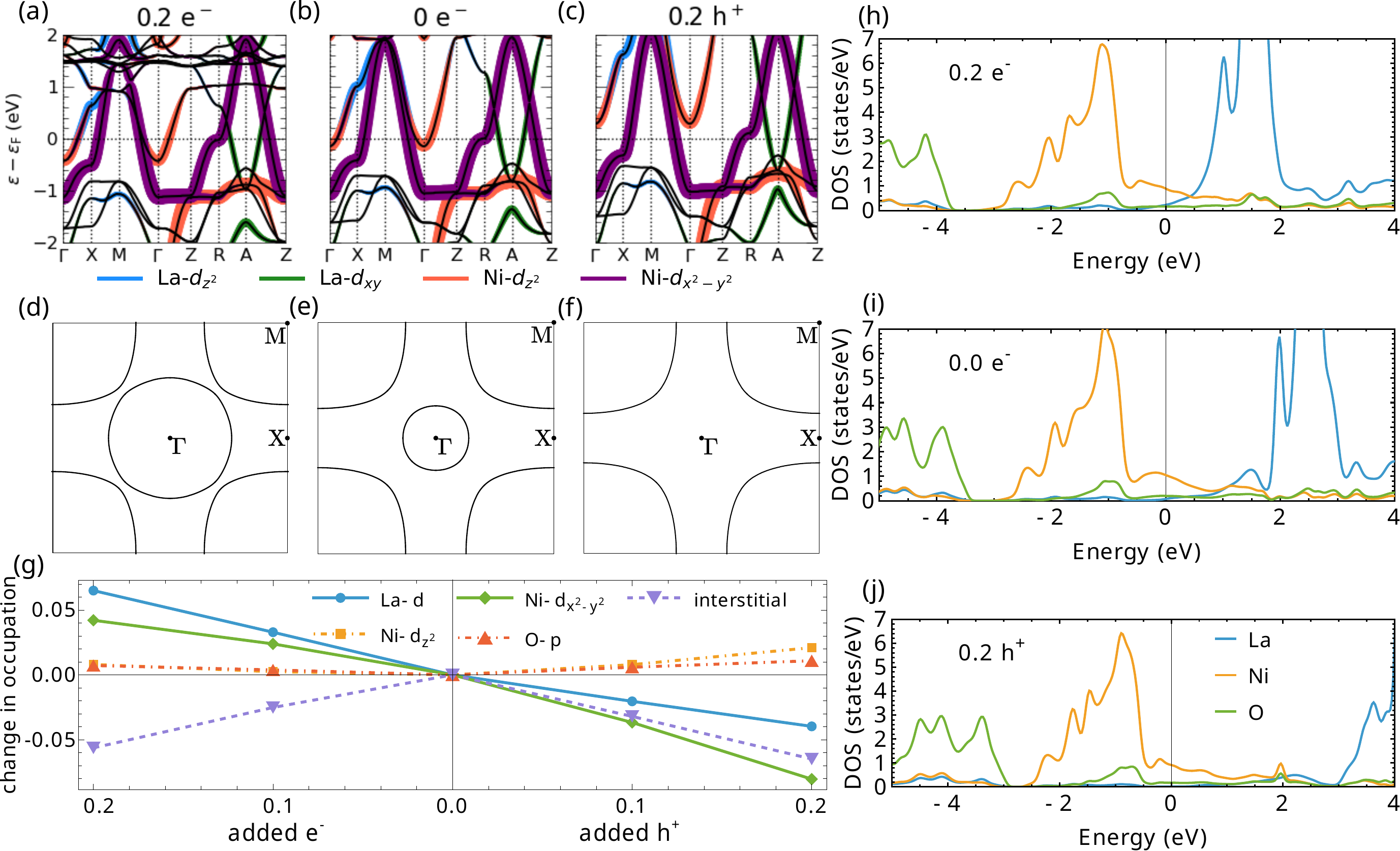}
    \caption{(a-c) Nonmagnetic DFT band structure of LaNiO$_2$ for selected electron/hole doping levels. Panels (d-f) show the corresponding Fermi surfaces for $k_z=0$ and (h-j) the Density of States (DOS). (g) Evolution of selected orbital projected occupations (in electrons) under electron and hole doping. Note that main changes take place in the interstitial, La-$d$ and Ni-$d_{x^2-y^2}$ orbitals. While the La-$d$ and Ni-$d_{x^2-y^2}$ occupations increase(decrease) upon electron(hole) doping, the interstitial occupation decreases in both doping regimes. 
    }
    \label{fig:DFT}
\end{figure*}

\section{Methods}

The effects of chemical doping in the infinite-layer nickelate LaNiO$_2$ were studied utilizing both supercell calculations and the virtual crystal approximation (VCA). The structural optimization of the supercells (with 12.5\% and 25\% doping levels) was performed using the VASP code and the projector-augmented-wave (PAW) approach   \cite{VASP,VASP-PAW} with the Perdew-Burke-Ernzerhof (PBE) \cite{perdew1996generalized} version of the generalized gradient approximation as the exchange-correlation functional. In VCA, the valence of La was fractionally varied according to the desired level of electron/hole doping. Further, in the VCA calculations the volume changes that electron/hole doping carry were linearly interpolated from the lattice constants of pure CeNiO$_2$ and SrNiO$_2$ for electron doping and hole doping, respectively. The optimized lattice parameters  are as follows $a=3.97$ \AA~ and $c=3.37$ \AA~ for LaNiO$_2$, $a=3.99$ \AA~ and $c=3.40$ \AA~ for SrNiO$_2$, and $a=3.92$ \AA~ and $c=3.35$ \AA~ for CeNiO$_2$. These values were also obtained using the VASP code and GGA-PBE. 
The electronic structure was subsequently analyzed in the nonmagnetic state at the DFT level using the all-electron full potential code {\sc wien2k} that employs an augmented plane wave plus local orbital (FP-APW+lo) basis set, also with GGA-PBE as the exchange-correlation functional \cite{blaha2020wien2k}. A $k$-mesh of $8\times 8 \times 10$ was used with $R_{MT}K_{max}=6$ and muffin tin radii ($R_{MT}$) of 2.5, 2.1, and 1.62 a.u., respectively, for La, Ni, and O.

For the spin-polarized DFT calculations aimed at understanding magnetic trends in the electron- and hole-doping regimes, an on-site
Coulomb repulsion $U$ was included using the LDA+$U$ method within the fully
localized limit (FLL) \cite{fll} in order to properly account
for correlation effects in the Ni($3d$) electrons. We used $U$-values ranging from
0 to 6 eV and a nonzero value of Hund’s coupling $J$=
0.8 eV. The following magnetic configurations
were considered: a) a ferromagnetic (FM) order, b) a C-type antiferromagnetic (AFM) order (checkerboard AFM planes coupled FM out-of-plane) and a G-type AFM order (checkerboard AFM planes coupled AFM).

DFT+DMFT calculations were performed using the TRIQS software package including the continuous-time quantum Monte Carlo impurity solver within the hybridization expansion \cite{TRIQS2015,CTHYB2016}. Localized Ni-$e_{g}$ ($d_{x^2-y^2}, d_{z^2})$ orbitals were constructed from the Kohn-Sham DFT band states using the TRIQS/DFTtools package, including all states within $\pm 10$ eV of the Fermi level~\cite{DFTTools2016}. The Kanamori parametrization of the Coulomb interaction was used. We employed $U=5$ eV and $J= 0.8$ eV \cite{Nowadnick2015Correlation,Chen2022} within the Held's double counting scheme, as defined in Ref. \cite{Held2007DC}:
\begin{equation}
    \Sigma_{dc} = \frac{U+(d-1)(U-2J)+(d-1)(U-3J)}{2d-1}\left(n-\frac{1}{2}\right)
\end{equation}
where $d$ is the total number of correlated orbitals and $n$ is the density in a given orbital. 
We fit our imaginary time Green's function to an order 40 Legendre function for stability \cite{PhysRevB.84.075145}. We used an inverse temperature of $\beta=40$ eV$^{-1}$ --roughly room temperature.

\section{Results}

\textit{DFT Nonmagnetic Electronic Structure.} Figure~\ref{fig:DFT} shows the DFT electronic
structure of LaNiO$_2$ upon both electron and hole doping. The results obtained using the VCA approximation are shown in the main text, while supercell calculations showing analogous trends are shown in Appendix~\ref{app:a}. The band structure of undoped LaNiO$_2$ (Fig. \ref{fig:DFT}(b)) has been intensively studied in the literature \cite{ES_112, arita, Liu2020, Thomale_PRB2020, Choi2020, Ryee2020,olevano20, Leonov2020,lechermann2020late,lechermann2020multi, Hu2019, Sakakibara, werner2020, eff_ham, Zhang2020, kitatani2020}. It portrays a single Ni-$d_{x^2-y^2}$ band crossing the Fermi level but with two extra electron pockets of La-$d_{xy}$ character (at A) and of La-$d_{z^2}$ character (at $\Gamma$). These additional rare-earth bands give rise to the self-doping effect that has been the subject of ample scrutiny.  While symmetry forbids hybridization of the Ni-$d_{x^2-y^2}$ band with either of these two bands, there is a large degree of hybridization between the La and Ni-$d_{z^2}$ states. 

Upon hole ($p$-type) doping (Fig.~\ref{fig:DFT}(c)), the most notable change is in the La-$d_{z^2}$ electron-pocket at the $\Gamma$ point, which completely vanishes by a doping level of $x$=0.2. The electron pocket at A of La-$d_{xy}$ character also shifts up in energy, together with the van Hove singularity at R that shifts away from the Fermi level with respect to the undoped case. Upon electron ($n$-type) doping (Fig. \ref{fig:DFT}(a)), both the La-$d_{z^2}$ and La-$d_{xy}$ electron pockets increase in size instead.  These changes can be better observed in the continuous evolution of the Fermi surface topology upon doping, as shown in Fig. \ref{fig:DFT}(d-f) for $k_z$=0 --the equivalent changes for $k_z$=1/2$\pi/c$ are shown in Appendix \ref{app:b}. Fig. \ref{fig:DFT}(h-j) shows the atom-resolved Density of States (DOS) for these different levels of doping. The primary features described above for the band structure are all visible: in particular, the upward shift of the La-$d$ states that move gradually to higher energies from the electron to the hole doped cases. Another sizable shift can be observed in the O-$p$ states that concomitantly shift up to higher energies when going from the electron to the hole-doping cases.

After understanding the most relevant changes in the electronic structure of LaNiO$_2$ upon doping, we now trace the carrier distribution of the dominant orbitals in the vicinity of the Fermi level upon both electron and hole doping (Figure \ref{fig:DFT}(g)). In particular, we monitor the change in occupation for the  Ni-$d_{x^2-y^2}$, Ni-$d_{z^2}$, La-$d$ ($d_{z^2}$+ $d_{xy}$), and interstitial states at different electron and hole doping levels. The first three orbitals represent the low-energy bands of infinite-layer nickelates. No relevant change in  O-$p$ occupation is observed, unlike cuprates, due to the large charge-transfer energy in infinite-layer nickelates. The extra carriers mainly go into the La-$d$ and Ni-$d_{x^2-y^2}$ orbitals, with an increase in occupation for $n$-type doping and a decrease in occupation in $p$-type doping. The $d_{z^2}$ orbital shows fairly small changes, consistent with its low weight near the Fermi level.  The interstitial charge decreases upon both electron and hole doping. These trends indicate that the missing/extra charge (corresponding to doped holes/electrons, respectively) is around the La, the Ni-site, and the middle of the Ni-Ni bond, where the interstitial state lies. Importantly, having larger electron pockets in the electron-doped side, does not effectively self-hole-dope the Ni-d$_{x^2-y^2}$ band with respect to half-filling. This change can be seen when only volume compression takes place (in analogy to pressure \cite{Meier2024,DiCataldo2024}), but once electrons are added they counteract for this effect in the Ni-$d_{x^2-y^2}$ orbital, with self-doped-holes going to the interstitial instead. Hence, while naively one could have expected electron and hole doping to give rise to the same effect: hole doping the $d_{x^2-y^2}$ orbital via added carriers (hole-doping) or via self-doping through the increase in size of the electron pockets (electron-doping), the picture is ultimately more complex resulting into a clear asymmetry in the $p$ and $n$-type doping sides of the phase diagram.

\begin{figure}
    \centering
    \includegraphics[width=0.9\linewidth]{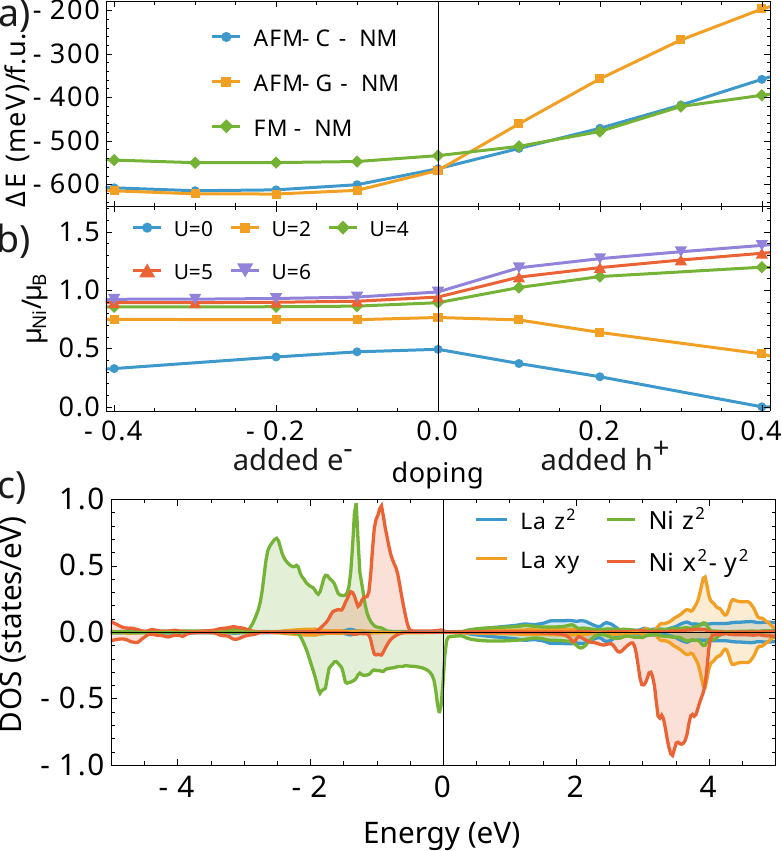}
    \caption{(a) Energy difference between different magnetic configurations (AFM-C, AFM-G, FM) with respect to a nonmagnetic (NM) state for $U=5, J=0.8$ eV. (b) Corresponding evolution of magnetic moments in the AFM-C state and different $U$, all with $J=0.8$ eV. Similar moment trends are obtained in a FM and AFM-C state. (c) Projected local spin-polarized DOS for the undoped C-type AFM state at $U=5$, $J=0.8$ eV. The DOS for a G-type state is very similar (as shown in Appendix \ref{app:c}). 
    }
    \label{fig:magnetic}
\end{figure}

\textit{Magnetic Trends in DFT: static correlations.} Antiferromagnetic spin fluctuations are considered a key ingredient in the cuprates and while antiferromagnetism is quickly suppressed upon hole-doping, the ordered antiferromagnetic state is much more robust upon electron-doping \cite{Armitage2010}. In infinite-layer
nickelates there is no experimental evidence for long-range antiferromagnetic order  even though strong magnetic excitations have been reported in the undoped compound \cite{rixs112}. From first-principles, both a C-type and a G-type antiferromagnetic (AFM) state are more stable (and closely competing in energy) than a nonmagnetic state for LaNiO$_2$  \cite{Kapeghian2020-wo}.

\begin{figure}
    \centering
    \includegraphics[width=\linewidth]{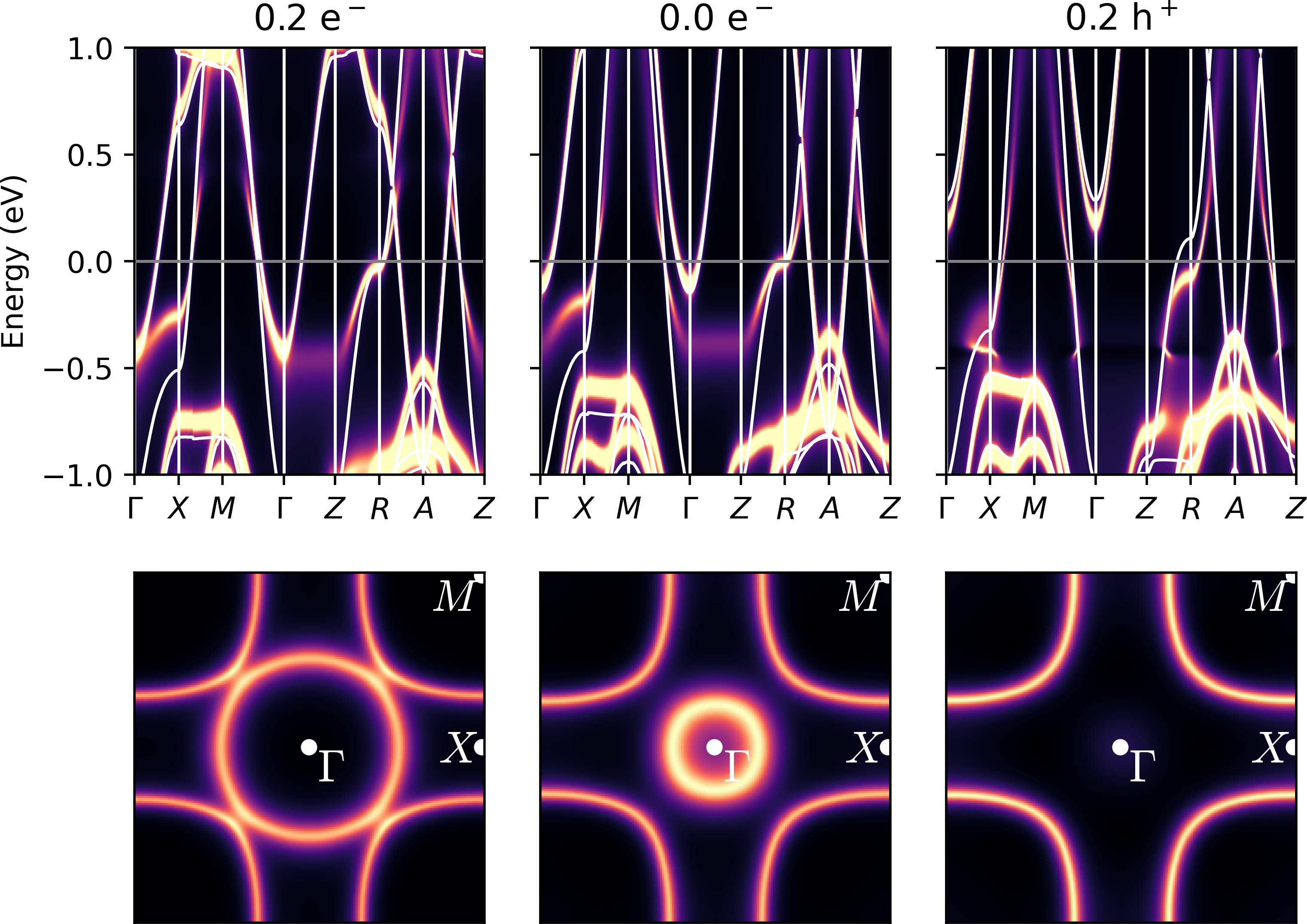}
    \caption{Momentum resolved spectral functions (top) and Fermi surfaces (bottom) at three levels of doping from DFT+DMFT calculations.\label{fig:Akw}}
\end{figure}

We study here the evolution of these two antiferromagnetic states upon hole and electron doping to better understand magnetic trends on both sides of the phase diagram. Figure \ref{fig:magnetic}(a) shows the changes in energy difference between these magnetic states (as well as a ferromagnetic one), with respect to the nonmagnetic state, as a function of effective doping. As in the nonmagnetic case, we used the VCA here to simulate chemical doping. There is a clear asymmetry as antiferromagnetism is quickly suppressed upon hole doping, while it remains the ground state for electron doping. For the chosen $U$= 5 eV, the C-type AFM and FM configurations turn out to be quasi-degenerate at $x$ = 0.2 (where superconductivity is conventionally observed in these materials), while the quasi-degeneracy of the C- and G-type AFM states obtained for the parent compound is preserved upon electron doping. The ground state trends remain similar for other choices of $U$ (see Appendix \ref{app:c}). The corresponding changes in the Ni magnetic moment are shown for the C-type AFM configuration for various choices of $U$ in Fig. \ref{fig:magnetic}(b). The results for the G-type AFM case are similar and shown in Appendix \ref{app:c}. 

The spin-polarized PDOS for the C-type AFM state is shown for one Ni site in Fig. \ref{fig:magnetic}(c) (the other site simply has the spin-up and spin-down DOS reversed due to the opposing orientation of the local moment). The PDOS shows that the Ni-$d$ states dominate around the Fermi level (analogous to the nonmagnetic description provided above). The clear asymmetry between electron and hole doping can be traced to the evolution of the magnetization -- that does not vary linearly with doping but changes asymmetrically instead.  Looking at the DOS, adding electrons fills only the minority-spin component, thereby suppressing magnetism, while removing electrons (hole-doping) does not empty the minority spin component, as one would expect for a linear trend, but instead depletes the majority Ni-$d$ component (see further details in Appendix \ref{app:c}). This leads to the nonlinear magnetization vs. doping behavior: a linear behavior would require the DOS for both spin components to have similar shapes across the entire energy range around the Fermi level.

\textit{Dynamical correlations.} The agreement found between VCA and supercell calculations at the DFT level confirms the reliability of the VCA method to study doped infinite-layer nickelates. Hence, we continue using VCA to study dynamical correlations in LaNiO$_2$. Figure \ref{fig:Akw} shows the momentum-resolved spectral functions along the same $k$-path used in our DFT calculations in Fig. \ref{fig:DFT}, as well as the corresponding ($k_z=0$) Fermi surfaces. It can be observed how the Ni-$d_{x^2-y^2}$ orbital crossing the Fermi energy is strongly renormalized compared to the DFT result (white lines). The electron pockets at the $\Gamma$ and A points essentially follow the DFT band structure, with the Fermi surfaces clearly showing the expansion or elimination of the $\Gamma$ pocket under electron and hole doping, respectively. As the size of the pockets changes upon doping, there is an overall change in $d_{x^2-y^2}$ bandwidth.

We next analyze the strength of electronic correlations via the orbital-resolved self-energy (Appendix~\ref{app:d}). Across all doping levels, the $d_{z^{2}}$ self-energy remains much smaller than the $d_{x^{2}-y^{2}}$, indicating weak correlations in the $d_{z^{2}}$ orbital. To quantify the doping dependence, we extract the mass enhancement  $m^*/m = 1 - \partial \im \Sigma(\omega_n\to0)/\partial\omega_n$ (see Table \ref{tab:mass_renorm}). While the $d_{z^{2}}$ mass enhancement remains independent of doping, the $d_{x^{2}-y^{2}}$ enhancement decreases under both hole and electron doping, with an almost twofold larger reduction on the electron-doped side. These results demonstrate that the low-energy correlation physics of infinite-layer nickelates lives on a single Ni-$d_{x^{2}-y^{2}}$ orbital that is weakly coupled to the charge-reservoirs.

\begin{table}[]
    \centering
    \begin{ruledtabular}
    \begin{tabular}{c|c|c}
        added $e^{-}$ & $d_{x^2-y^2}$ & $d_{z^2}$ \\\hline
        -0.20 & 2.75 & 1.21 \\
        0.00 & 2.91 & 1.23 \\
        0.20 & 2.51 & 1.24 
    \end{tabular}
    \end{ruledtabular}
    \caption{Mass enhancement, $m^*/m$, for the two $e_g$ orbitals at three levels of doping. There is a clear change in mass enhancement for the $x^2-y^2$ orbital, but not the $z^2$ orbital. }
    \label{tab:mass_renorm}
\end{table}

\section{Conclusions}

 Our results show that there is a clear asymmetry in the electronic structure and magnetic properties obtained upon electron and hole doping in infinite-layer nickelates. Electron doping increases the size of the ``self-doping'' electron pockets from the $R(5d)$ states, however, this does not effectively increase the hole doping effect of the Ni-$d_{x^2-y^2}$ band with respect to half-filling (as naively expected). We also predict that the $e$-doped infinite-layer nickelates should show stronger antiferromagnetic interactions. The clear differences we observe between $p$ and $n$-type doping motivate electron-doped infinite-layer nickelates as an interesting experimental pursuit. If topotatic reduction of Ce or Th doped samples is difflcult, gating can be another strategy to access this regime. While we do not study superconducting trends, the likely scenario in light of our results (in combination with the reports of superconductivity at zero doping) is the possibility of an extended superconducting dome on the hole-doped side of the phase diagram. Carefully exploring lightly-hole-doped samples could help shed some light on this issue.

\section{Acknowledgements}
F.B. acknowledges support from the European Union - Next Generation EU, within the PRIN 2022 call, project SUBLI  contract n.~2022M3WXE7 and project PRIN 2022 TOTEM, grant n.~F53D23001080006, funded by Italian Ministry of University and Research (MUR). ASB, EDR and YZ acknowledge NSF grant No. DMR-2045826 and the
ASU research computing center for HPC resources. The Flatiron Institute is a division of the Simons Foundation.

\appendix

\section{Supercell calculations}\label{app:a}
\begin{figure}
    \centering
    \includegraphics[width=0.9\linewidth]{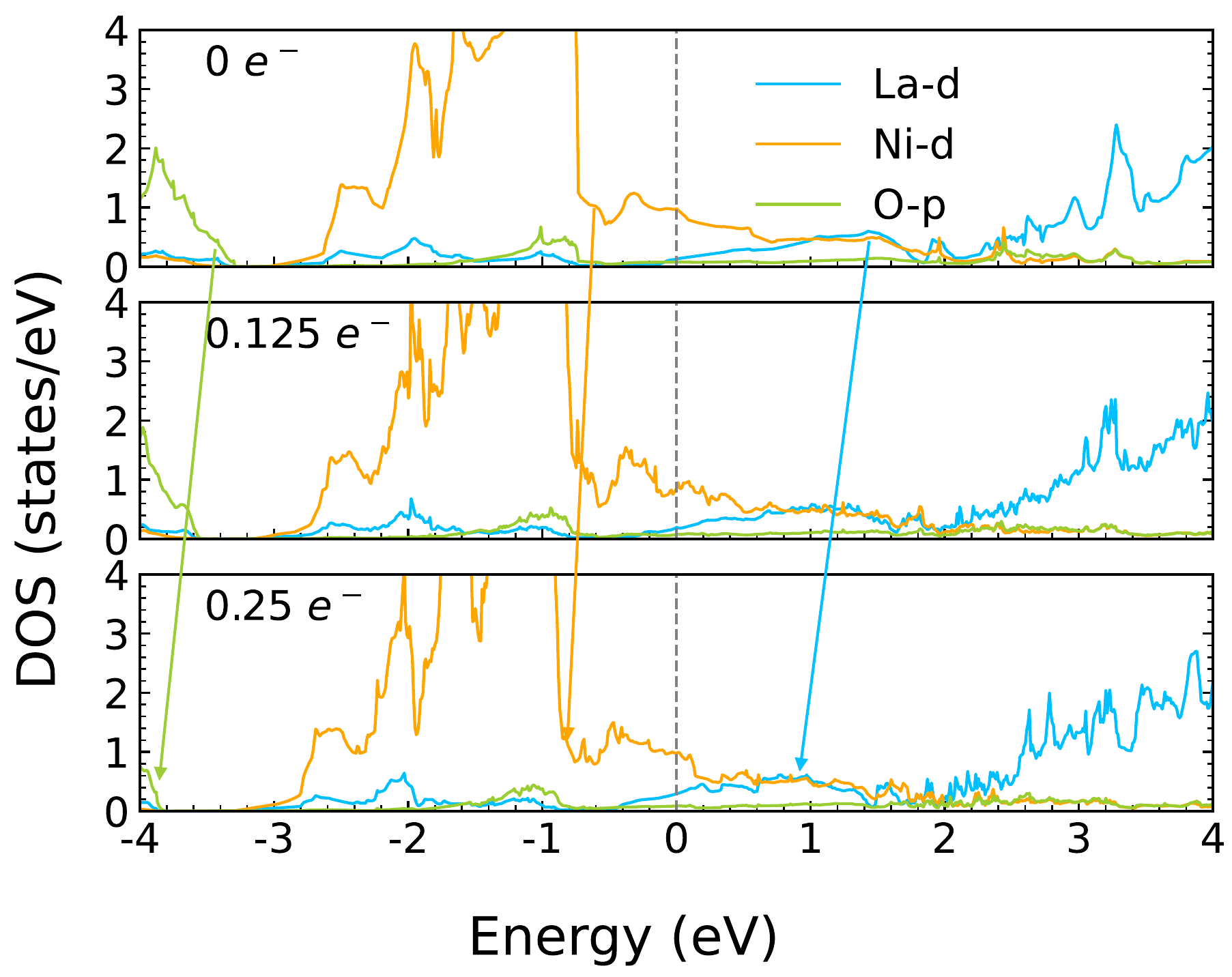}
    \caption{Atom and orbital-resolved DOS for La$_{1-x}$Ce$_x$NiO$_2$ in supercell calculations. Arrows have been added to illustrate the shift in the DOS with doping.}
    \label{fig:supercell}
\end{figure}

In order to assess the validity of the VCA method we performed additional supercell calcultations for La$_{1-x}$Ce$_x$NiO$_2$, whose resulting DOS is shown in Fig. \ref{fig:supercell} for different levels of doping. The same shifts as those in the VCA calculations can be observed: both the O-$p$ and La-$d$ states shift down to lower energies upon increasing electron doping. Calculations upon hole doping with Sr were already performed by us (see Ref. \cite{krishna}), also showing trends that agree with the VCA ones. The supercell results hence provide support for our use of VCA in the results described in the main text.

\section{Fermi surface evolution}\label{app:b}

\begin{figure}
    \centering
    \includegraphics[width=0.83\linewidth]{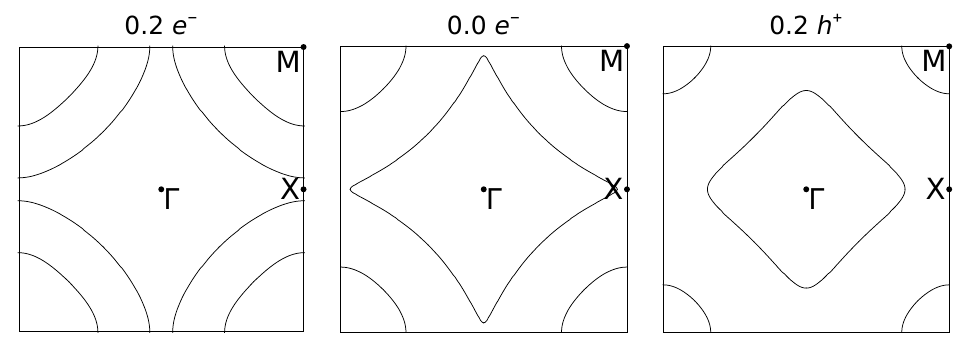}
    \caption{Fermi surfaces for $k_z=\pm\frac{\pi}{c}$ at three levels of doping for LaNiO$_2$.}
    \label{fig:kzhalf}
\end{figure}

Fig. \ref{fig:kzhalf} shows the evolution of the Fermi surface at $k_z=\pm\pi/c$ upon doping for LaNiO$_2$ within VCA. Changes in the Ni-$d_{x^2-y^2}$-dominated Fermi surface from hole- to electron-like can be observed as the van Hove singularity is crossed. The La-$d_{xy}$-pocket at the zone corners gets reduced from the electron- to hole-doping cases. Aside from the van Hove singularity crossing, these are similar trends to those of the $k_z=0$ Fermi surfaces.

\section{Additional details of the DFT magnetic calculations}\label{app:c}

The following figures illustrate additional results of the spin-polarized DFT calculations to demonstrate that our conclusions do not depend on the precise value of Hubbard $U$ or the AFM order described in the main text.

\begin{figure}
    \centering
    \includegraphics[width=0.85\linewidth]{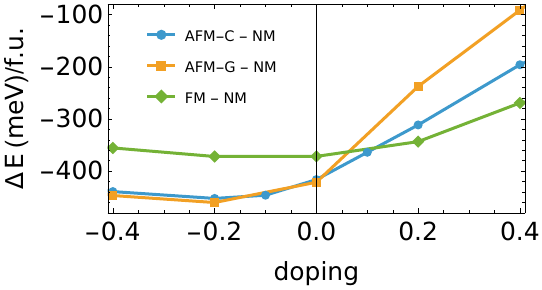}
    \caption{Energy difference between different magnetic configurations (AFM-C, AFM-G, FM) with respect to a nonmagnetic (NM) state for $U$ = 4, $J$ = 0.8 eV}
    \label{fig:U4-energetics}
\end{figure}

Fig. \ref{fig:U4-energetics} shows the change in energy difference between the three magnetic states discussed in the main text with respect to the nonmagnetic state as a function of doping for a different $U$ value of 4 eV. The overall trends are very similar to those for $U=5$ eV depicted in Fig. \ref{fig:magnetic}(a) of the main text. The primary change is a larger energy difference between the FM and AFM-C state and a small shift in the relative energies between these two states with respect to doping in the hole-doped region.

\begin{figure}
    \centering
    \includegraphics[width=0.9\linewidth]{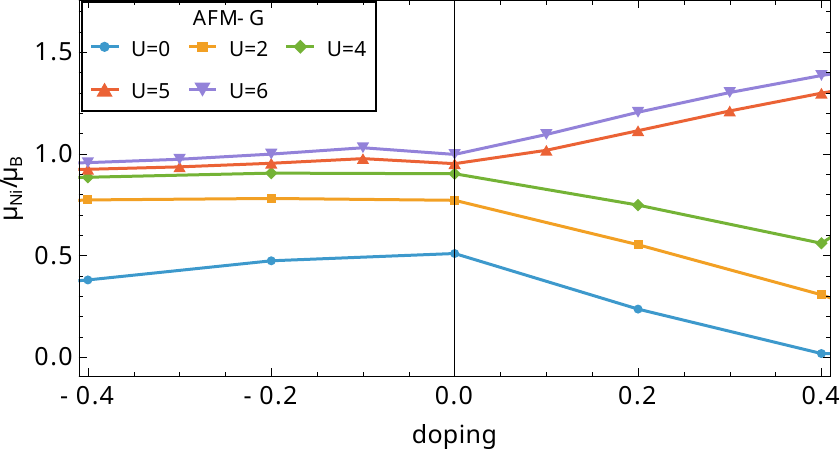}
    \caption{Evolution of magnetic moments for various values of $U$ and $J$ in the AFM-G configuration.}
    \label{fig:additionalMoments}
\end{figure}

Fig. \ref{fig:additionalMoments} shows the evolution of the magnetic moments for various values of $U$ for different doping levels for the AFM-G state. The trends are very similar to the AFM-C ones shown in Fig. \ref{fig:magnetic}(b) of the main text. The sharper changes are obtained above $U=4$ with the onset of a spin-state transition from low to high-spin. The moments can be seen to increase with $U$, as expected.

\begin{figure}
    \centering
    \includegraphics[width=0.8\linewidth]{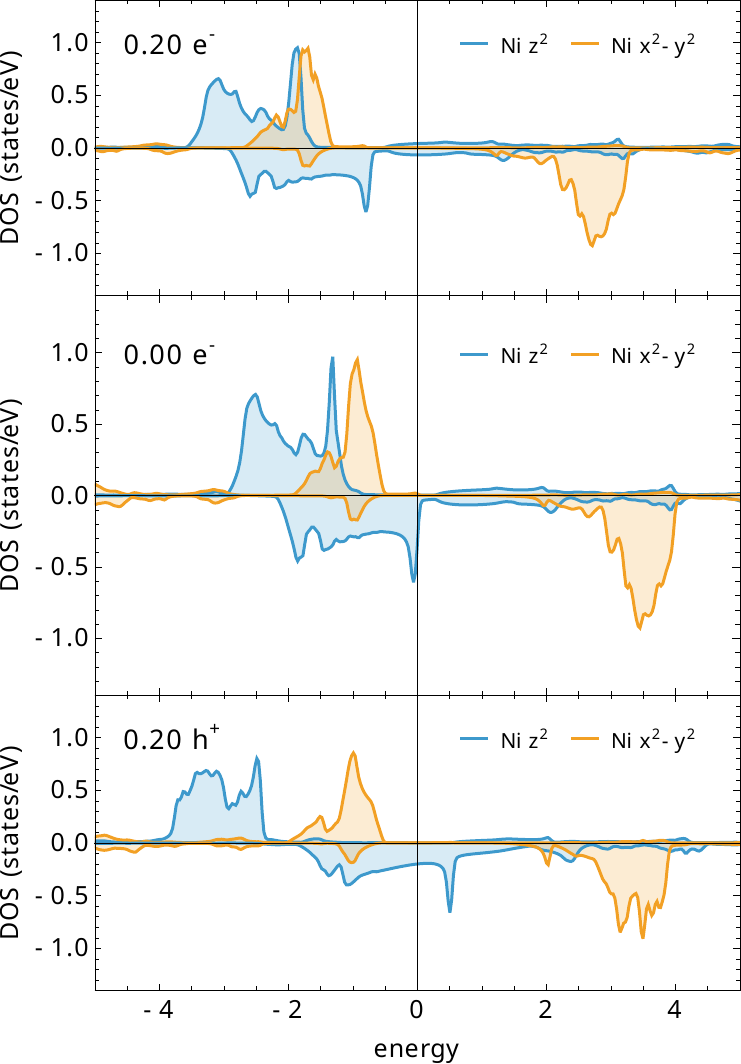}
    \caption{Ni spin-polarized PDOS for the AFM-C configuration at three levels of doping.}
    \label{fig:DOS_doped}
\end{figure}

The spin-projected DOS for the Ni $e_g$ orbitals at three levels of doping are shown in Fig. \ref{fig:DOS_doped}. The asymmetry between electron and hole-doping described in the main text can be observed. 

\section{Self-energies}\label{app:d}
\begin{figure}
    \centering
    \includegraphics[width=\linewidth]{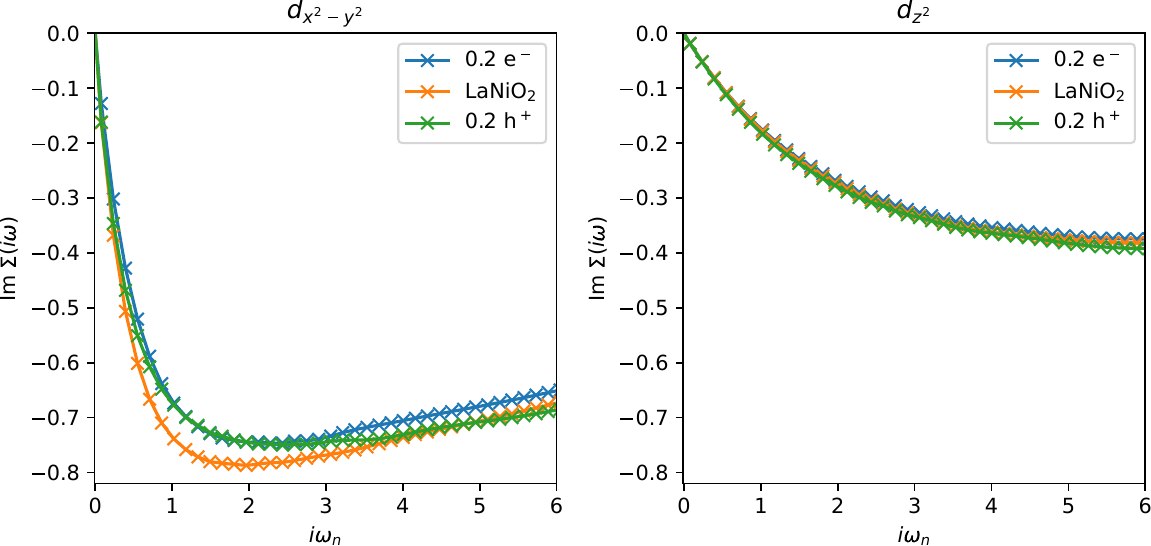}
    \caption{Self-energies of both $d_{x^2-y^2}$ and $d_{z^2}$ orbitals across three doping levels.}
    \label{fig:self-energies}
\end{figure}

The self-energies for both $e_g$ orbitals across the three doping levels is shown in Fig. \ref{fig:self-energies}. These expand on the trends from Table \ref{tab:mass_renorm} of the mass enhancements, showing the steepest slope near zero in the undoped case and electron doping reducing the slope more than hole doping. The $d_{z^2}$ orbital remains weakly correlated with minimal change under doping, again emphasizing that the $d_{x^2-y^2}$ orbital is the one of primary interest for the low-energy physics in these systems.

\bibliographystyle{apsrev4-1}
\bibliography{LaNiO2}

\end{document}